# Fusion of convolution neural network, support vector machine and Sobel filter for accurate detection of COVID-19 patients using X-ray images

Danial Sharifrazi[1], Roohallah Alizadehsani[2], Mohamad Roshanzamir[3], Javad Hassannataj Joloudari[4], Afshin Shoeibi[5,6,*], Mahboobeh Jafari[7], Sadiq Hussain[8], Zahra Alizadeh Sani[9,10], Fereshteh Hasanzadeh[10], Fahime Khozeimeh[2], Abbas Khosravi[2], Saeid Nahavandi[2], Maryam Panahiazar[11], Assef Zare[12], Sheikh Mohammed Shariful Islam[13,14, 15], U Rajendra Acharya[16,17,18]

[1] Department of Computer Engineering, School of Technical and Engineering, Shiraz Branch, Islamic Azad University, Shiraz, Iran
[2] Institute for Intelligent Systems Research and Innovations (IISRI), Deakin University, Geelong, Australia
[3] Department of Engineering, Fasa Branch, Islamic Azad University, Post Box No 364, Fasa, Fars 7461789818, Iran.
[4] Department of Computer Engineering, Faculty of Engineering, University of Birjand, Birjand, Iran
[5] Computer Engineering Department, Ferdowsi University of Mashhad, Mashhad, Iran.
[6] Faculty of Electrical and Computer Engineering, Biomedical Data Acquisition Lab, K. N. Toosi University of Technology, Tehran, Iran.
[7] Electrical and Computer Engineering Faculty, Semnan University, Semnan, Iran
[8] System Administrator, Dibrugarh University, Assam 786004, India
[9] Rajaie Cardiovascular Medical and Research Center, Iran University of Medical Sciences, Tehran, Iran
[10] Omid hospital, Iran University of Medical Sciences, Tehran, Iran
[11] Institute for Computational Health Sciences, University of California, San Francisco, USA.
[12] Faculty of Electrical Engineering, Gonabad Branch, Islamic Azad University, Gonabad, Iran
[13] Institute for Physical Activity and Nutrition, Deakin University, Melbourne, Australia
[14] Cardiovascular Division, The George Institute for Global Health, Australia
[15] Sydney Medical School, University of Sydney, Australia
[16] Department of Electronics and Computer Engineering, Ngee Ann Polytechnic, Singapore,
[17] Department of Biomedical Engineering, School of Science and Technology, Singapore University of Social Sciences, Singapore,
[18] Department of Bioinformatics and Medical Engineering, Asia University, Taiwan.
* Corresponding author: Afshin Shoeibi, Computer Engineering Department, Ferdowsi University of Mashhad, Mashhad, Iran.
E-mail: afshin.shoeibi@gmail.com

## Abstract

The coronavirus (COVID-19) is currently the most common contagious disease which is prevalent all over the world. The main challenge of this disease is the primary diagnosis to prevent secondary infections and its spread from one person to another. Therefore, it is essential to use an automatic diagnosis system along with clinical procedures for the rapid diagnosis of COVID-19 to prevent its spread. Artificial intelligence techniques using computed tomography (CT) images of the lungs and chest radiography have the potential to obtain high diagnostic performance for Covid-19 diagnosis. In this study, a fusion of convolutional neural network (CNN), support vector machine (SVM), and Sobel filter is proposed to detect COVID-19 using X-ray images. A new X-ray image dataset was collected and subjected to high pass filter using a Sobel filter to obtain the edges of the images. Then these images are fed to CNN deep learning model followed by SVM classifier with ten-fold cross validation strategy. This method is designed so that it can learn with not many data. Our results show that the proposed CNN-SVM with Sobel filtering (CNN-SVM+Sobel) achieved the highest classification accuracy of 99.02% in accurate detection of COVID-19. It showed that using Sobel filter can improve the performance of CNN. Unlike most of the other researches, this method does not use a pre-trained network. We have also validated our developed model using *six* public databases and obtained the highest performance. Hence, our developed model is ready for clinical application.

# 1. Introduction

Coronavirus disease 2019 (COVID-19) has been spreading unprecedentedly across the globe from the beginning of 2020. The clinical characteristics of COVID-19 include respiratory symptoms, fever, cough, dyspnea, pneumonia, and fatigue during early stages [1], [2], [3]. The COVID-19 also affects the cardiovascular and respiratory systems and may lead to multiple organ failure or acute respiratory distress in critical cases and is highly contagious [3-7]. Therefore, COVID-19 infections are a crucial healthcare challenge around the world and has become a global threat [8].

The World Health Organization (WHO) declared the outbreak a "public health emergency of international concern" on 30[th] January 2020. Reverse-transcription polymerase chain reaction (RT-PCR) is generally used to confirm the incidence of COVID-19. But the sensitivity of RT-PCR is not high enough for the early recognition of suspected patients [9]. Recently, deep learning techniques have exhibited great success in the image processing domain, especially medical images, due to its potential of feature extraction [10]. Deep learning is used to discriminate and detect viral and bacterial pneumonia in pediatric chest radiographs [11]. Chest X-ray is found to be effective in the early diagnosis and screening of COVID-19 [12]. This non-invasive imaging modality can help to detect specific characteristic manifestations in the lung related to the COVID-19. Hence, radiography examination may be utilized as a primary tool for COVID-19 screening in epidemic areas. Several imaging features can be extracted from the chest X-ray [13, 14]. In heavily-affected areas and resource-constrained areas, chest X-ray imaging can be beneficial for COVID-19 screening [15]. There are various advantages related to this imaging modality, such as rapid triaging, availability, accessibility, and portability [15]. It is cheap and can be made available in most of the clinical settings, even in low-income countries. One of the bottlenecks of the system is that expert radiologists are required to interpret the radiography images. As such, computer-aided diagnostic systems (CAD) can help the radiologists to detect COVID-19 cases accurately and rapidly. There are few deep learning-based techniques proposed for such automated detection using X-ray radiographs [15-25].

The main contributions of this work are listed as follows:

- New private database collected by the authors is used.
- Data augmentation is performed.
- Proposed model is tested using six public databases and the results are found to be better than most of the existing state of the art methods.
- Sobel filter is found to improve the performance of CNN.
- Obtained highest classification performance for all databases.

Nowadays, machine learning (ML) methods are widely used for Covid-19. These methods can improve the diagnosis accuracy of clinicians. However, there are few limitations in these methods. For example, feature extraction is a challenging step in almost all ML methods. So, automatic feature extraction is a great improvement in this field. Among the different ML methods, deep learning (DL) can solve this challenge. It can do feature extraction automatically. In addition, when there are large amount of data, its performance is better than other ML methods. Consequently, nowadays DL is used to diagnose different diseases [26-32] such as COVID-19 [25]. An overview of the works done on automated detection of COVID-19 using DL is presented in Table 1. In this table, the recently published DL works on COVID-19 detection using X-ray and CT scan images are listed. However, almost all of them used pre-trained networks using public databases.

Table 1: Summary of works done on automated detection of COVID-19 using DL techniques with X-ray and CT images.

| Study | Modality | Number of Cases (or Images) | Network |
|---|---|---|---|
| Wang et al. [14] | X-ray | 13,975 images | Deep CNN |
| Hall et al. [15] | X-ray | 455 images | VGG-16 and ResNet-50 |
| Farooq et al. [16] | X-ray | 5941 images | ResNet-50 |
| Hemdan et al. [18] | X-ray | 50 images | DesnseNet, VGG16, MobileNet v2.0 etc. |
| Abbas et al. [19] | X-ray | 196 images | CNN with transfer learning |
| Minaee et al. [20] | X-ray | 5000 images | DenseNet-121, SqueezeNet, ResNet50, ResNet18 |
| Zhang et al. [21] | X-ray | 213 images | ResNet, EfficientNet |
| Apostolopoulos et al. [23] | X-ray | 3905 images | MobileNet v2.0 |
| Narin et al. [24] | X-ray | 100 images | InceptionResNetV2, InceptionV3, ResNet50 |
| Luz et al. [13] | X-ray | 13, 800 images | EfficientNet |
| Brunese et al. [33] | X-ray | 6,523 images | VGG-16 and transfer learning |
| Ozturk et al. [34] | X-ray | Two publically available databases were used where images were updated regularly. | Darknet-19 |
| Khan et al. [35] | X-ray | 1251 images | CNN |
| Silva et al. [36] | CT scans | 2482 images | A slice voting-based approach extending the Efficient Net Family of deep artificial neural networks |
| Luz et al. [13] | X-ray | 13, 800 images | Efficient Net |
| Ozturk et al. [34] | X-ray | Two publically available databases were used where images were updated regularly. | Darknet-19 |
| Khan et al. [35] | X-ray | 1251 images | CNN |
| Haghanifar et al. [37] | X-ray | 7,700 images | DenseNet-121 U-Net |
| Oh et al. [38] | X-ray | 502 images | DenseNet U-Net |
| Tartaglione et al. [39] | X-ray | 5 different databases | ResNet |
| Rahimzadeh et al. [40] | X-ray | 11302 images | Xception and ResNet50V2 |
| Jamil et al. [41] | X-ray | 14150 images | Deep CNN |
| Horry et al. [42] | X-ray | 60,798 images | VGG, Inception, Xception, and Resnet |
| Elasnaoui et al. [43] | X-ray And CT | 6087 images | inception_Resnet_V2 and Densnet201 |
| Ardakani et al. [44] | CT | 1020 | ResNet-101, ResNet-50, ResNet-18, GoogleNet, SqueezeNet, VGG-19, AlexNet |

This paper is organized as follows. The computer aided diagnosis (CAD) based on the proposed deep learning to detect COVID-19 is described in Section 2. The results obtained is presented and discussed in Section 3. Finally, the paper concludes with brief summary in Section 4.

## 2. CADS Based COVID-19 Diagnosis Using 2D-CNN

Nowadays, many CAD systems have been developed using deep learning techniques to detect various diseases, including COVID-19, have attracted the attention of many researchers. The CAD based deep learning methods require huge database to yield highest performance. This paper proposes a novel 2D-CNN architecture to detect COVID-19 using X-ray images. The 2D-CNN with a number of convolutional layers, max-pooling, and fully connected (FC) layers are used. In our methodology, support vector machine (SVM) is used instead of the sigmoid activation function in fully connected layers to obtain highest classification performance. The proposed CAD system is shown in Figure 1.

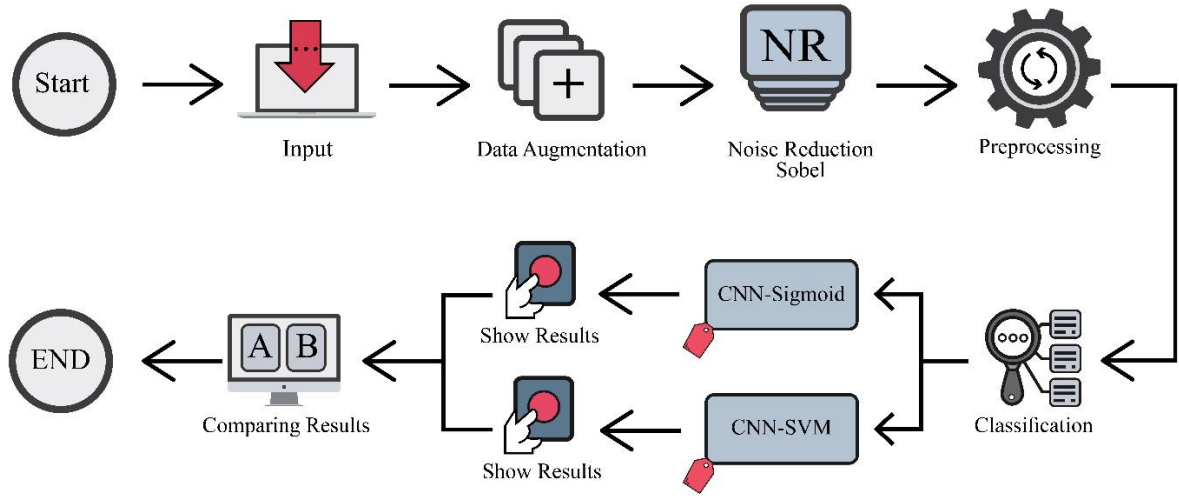

Figure 1: Proposed methodology used for the automated detection of COVID-19 patients using X-ray images.

As shown in Figure 1, X-ray images are first applied to the network. Then, the data augmentation technique is adopted to prevent the overfitting and increase the number of input data. Then during pre-processing stage, image resizing and normalization of input images are done. Then the preprocessed images are fed to the convolutional layers of the proposed 2D-CNN network to extract the features. Then, the classification operation is accomplished by FC layers by two methods: (i) sigmoid and (ii) SVM separately.

2.1. X-ray database

In this study, 333 chest X-ray images comprising of 77 images of COVID-19 patients and 256 images of normal subjects were recorded at Omid Hospital in Tehran. They are collected from February 2020 to April 2020. The mean and standard deviation of their age are 49.5±18.5 years old. 55% of cases are female. Three radiologists checked each image and determined whether a case has Covid-19 or not. Ethical approval of these data was also obtained. Some examples of these data can be seen in Figure 2. They show the typical X-ray images of normal and COVID-19 patients.

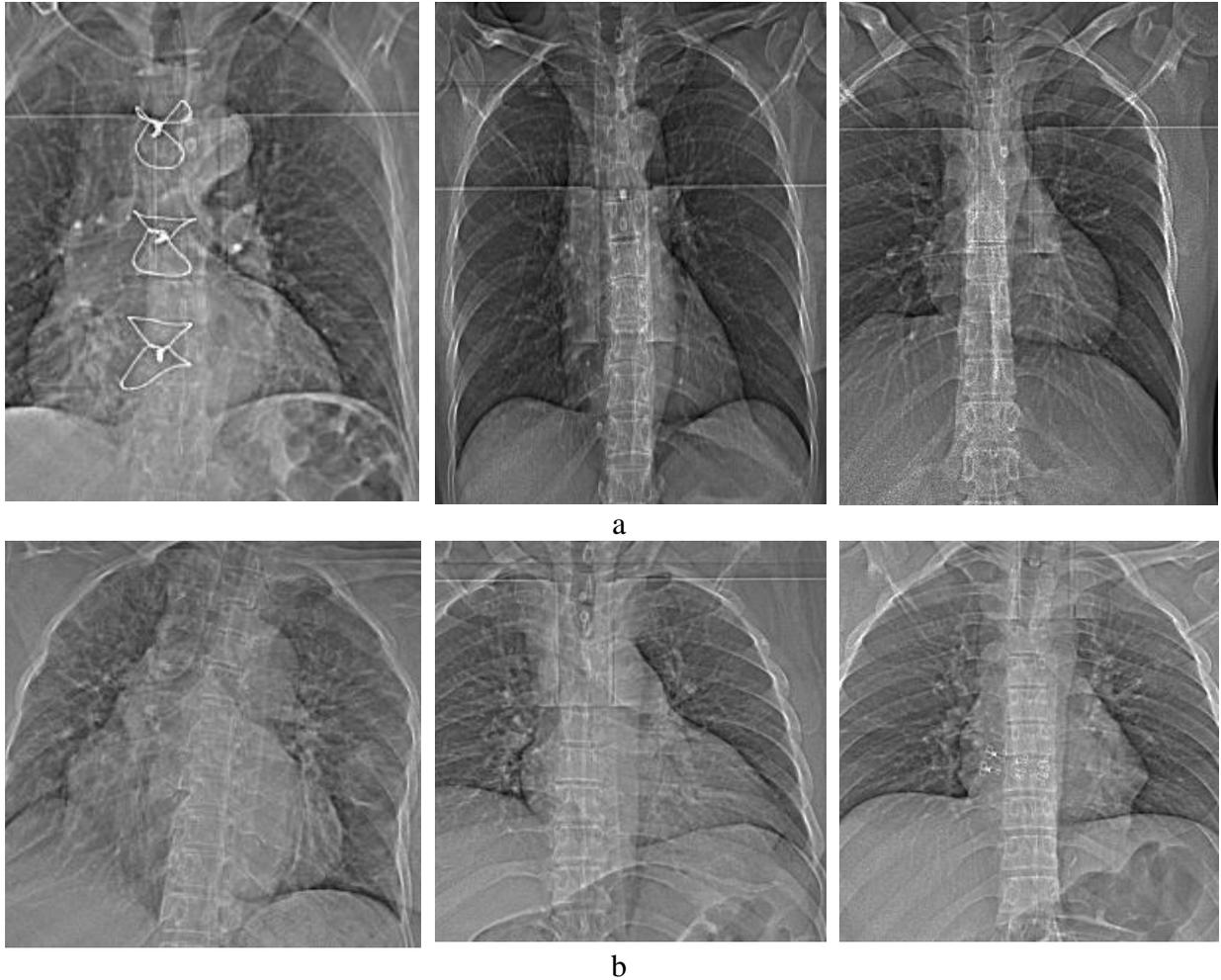

Figure 2. Sample X-ray images: a) healthy subjects and b) COVID-19 patients.

## 2.2. CNN architecture

Nowadays, 2D-CNN networks are employed in many medical applications, including the diagnosis of COVID-19 using X-ray images [45]. These deep learning networks consists of *three* main layers, convolutional layers, pooling, and fully connected (FC) layers [28]. The convolutional layers are responsible for extracting features from images. Max-pooling layers are often applied to reduce the features in CNN architectures. The last part of 2D-CNN is FC, and in the previous layer, there is an activation function that is responsible for classification. Usually, the Softmax function is employed. The Sigmoid activation function has been proved to perform efficiently in binary classification problems in this deep learning architecture. The support vector machine (SVM) is another procedure that can be applied in 2D-CNN instead of Sigmoid to obtain favorable results.

In this work, at first, the number of data is increased using data augmentation algorithm. Data augmentation is done by using width shift range, height shift range, and rotation techniques. Using this method, the data is increased from 333 to 1332 images. Then, a 2D-CNN with sigmoid activation function is used to classify X-ray images. In addition, binary SVM is also used in the 2D-CNN network for classification. The hinge error function is used to obtain best results when using SVM in 2D-CNN. More details about the proposed 2D-CNN architecture is shown in Table 2 and Figure 3.

Table 2: Details of parameters used in the proposed CNN architecture.

| Number of Kernels related to first and second connection | Size of the convolution kernels | Size of the max pooling kernels | Number of neurons in the Fully Connected layer | Number of neurons in the output layer | Size of the Dropout layer | Number of batch size | Number of epochs | Value of validation data | Optimizer function | Activator function | Loss function for CNN+Sigmoid | Loss function for CNN+SVM | SVM function kernel | Output layer classifiers |
|---|---|---|---|---|---|---|---|---|---|---|---|---|---|---|
| 128 and 256 | 3*3 | 2*2 | 64, 32 and 16 | 2(health and sick) | 0.2 | 32 | 100 | 0.3 and 0.2 | Adam | ReLU | binary cross entropy | Hinge | Linear | Sigmoid and SVM |

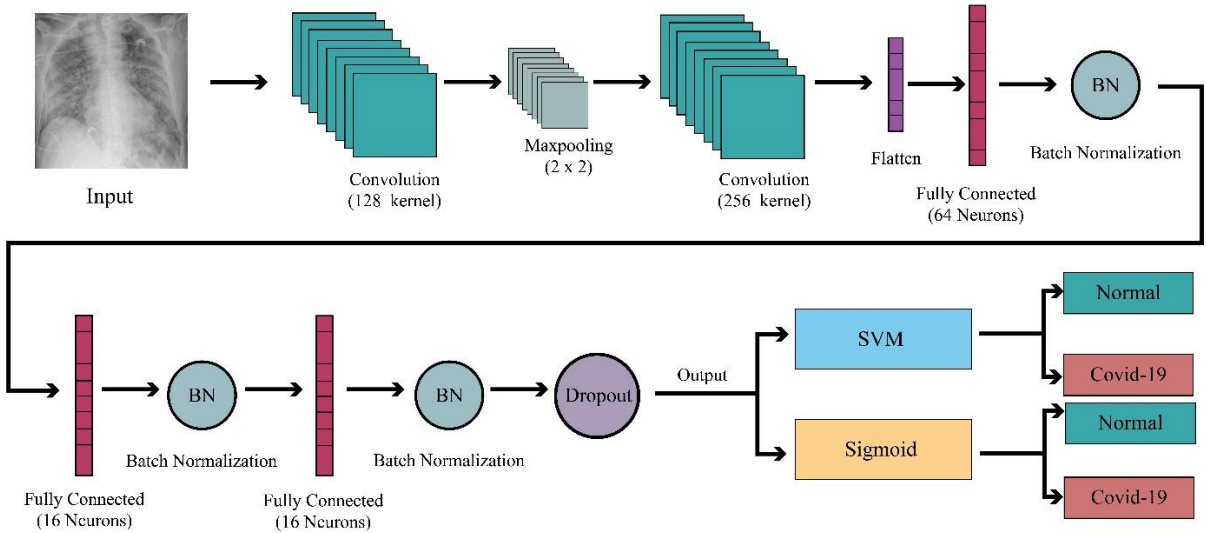

Figure 3: Proposed CNN architecture for the automated detection of COVID-19 patients using X-ray images.

2.3. Performance measures

In this study, to evaluate the performance of proposed methods, various evaluation metrics have been used and they are given below:

$$\text{Accuracy} = \frac{TP + TN}{FP + FN + TP + TN} \qquad (1)$$

$$\text{Sensitivity} = \frac{TP}{TP + FN} \qquad (2)$$

$$\text{Precision} = \frac{\text{TP}}{\text{TP} + \text{FP}} \tag{3}$$

$$\text{F1} - \text{Score} = \frac{2\text{TP}}{2\text{TP} + \text{FP} + \text{FN}} \tag{4}$$

$$\text{Specificity} = \frac{\text{TN}}{\text{TN} + \text{FP}} \tag{5}$$

In these equations, true positive (TP) is the correct classification of positive class. False-negative (FN) is the incorrect prediction of the positive case. True negative (TN) is the correct classification of the samples in the negative class. False-positive (FP) is the incorrect prediction of the negative case. In this work, positive class is symptom of COVID-19 and normal class is negative class.

## 3. Results and Discussion

In this section, the results of our proposed CNN-SVM and CNN-Sigmoid methods and its combination with Sobel filter are provided. All simulations are done using Keras library have been conducted with back-end TensorFlow. The COVID-19 X-ray images database is obtained from Omid Hospital, Tehran, Iran. In this work, total number of 1332 (total images number is 333, which is increased to 1332 after the data augmentation operation) images are used. The results are obtained in two modes: (i) CNN network with sigmoid output layer and (ii) CNN network with SVM output layer with 10-fold cross-validation strategy. In order to validate the proposed method, we have tested with another public database named as augmented COVID-19 X-ray images database [46]. The experiment results are presented in Figures 8 to 15.

Figure 8 illustrates the results obtained using private database with CNN-sigmoid method with 10-fold cross-validation. Figures 9 to 11, shows the private database results obtained by applying CNN-SVM, CNN-sigmoid with Sobel operator, and CNN-SVM with Sobel operator, respectively with 10-fold cross-validation.

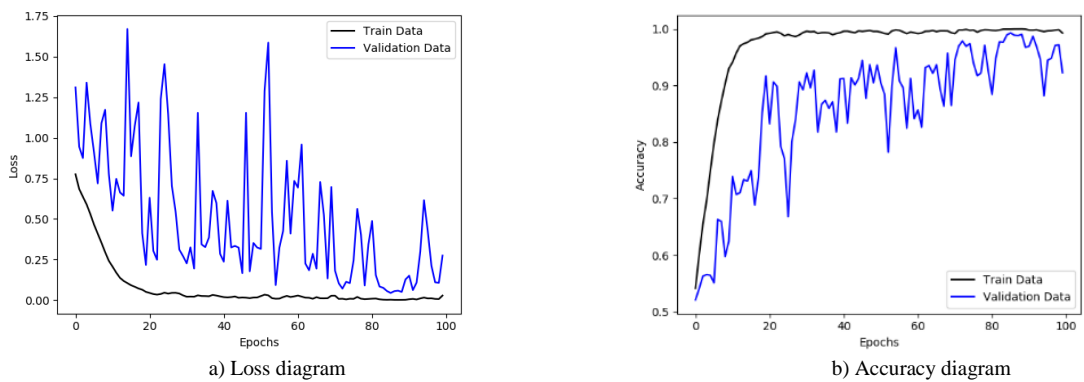

a) Loss diagram

b) Accuracy diagram

Figure 8: Performance metrics of CNN-sigmoid method using private database: (a) loss function curve, and b) accuracy curve with 10-fold cross-validation strategy.

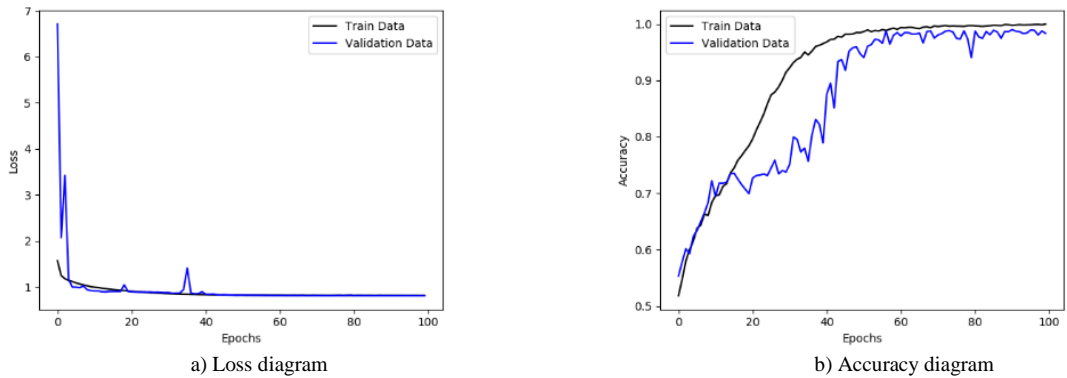

a) Loss diagram           b) Accuracy diagram

Figure 9: Performance metrics of CNN-SVM method using private database: (a) loss function curve, and b) accuracy curve with 10-fold cross-validation strategy.

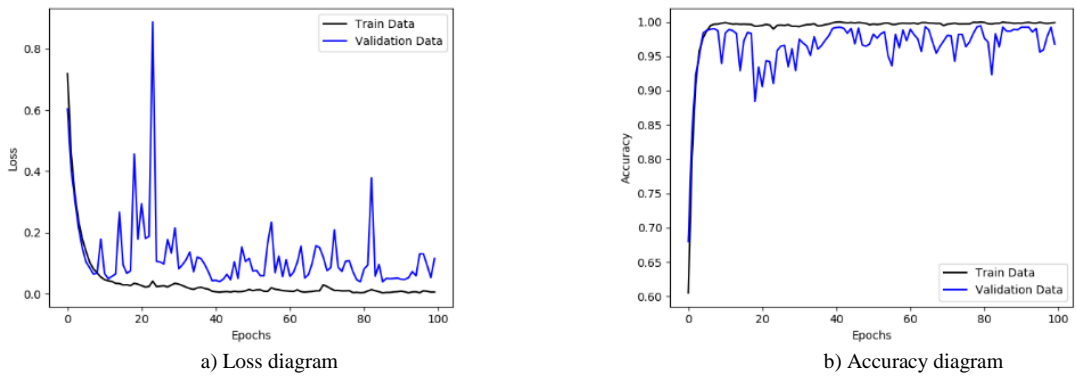

a) Loss diagram           b) Accuracy diagram

Figure 10: Performance metrics of CNN-sigmoid with Sobel operator method using private database: (a) loss function curve, and b) accuracy curve with 10-fold cross-validation strategy.

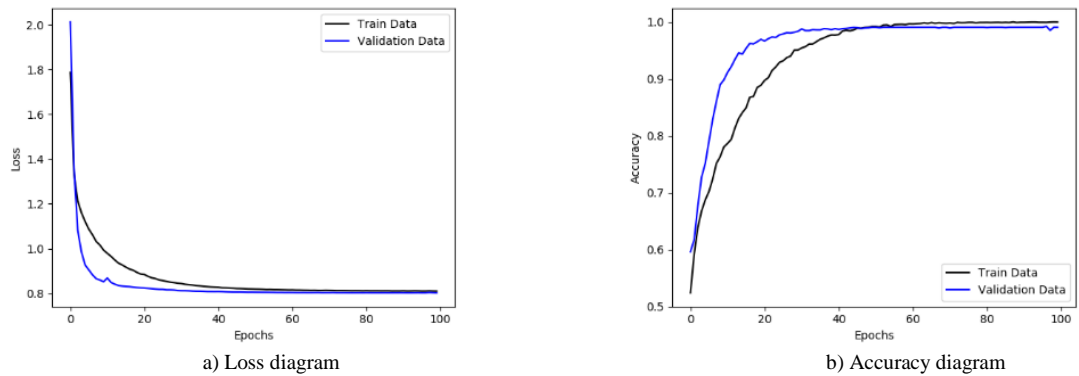

a) Loss diagram           b) Accuracy diagram

Figure 11: Performance metrics of CNN-SVM with Sobel operator method using private database: (a) loss function curve, and b) accuracy curve with 10-fold cross-validation strategy.

Figures 12 to 15, show the results obtained by applying CNN-Sigmoid, CNN-SVM, CNN-sigmoid with Sobel, and CNN-SVM with Sobel operator respectively with 10-fold cross-validation strategy using augmented COVID-19 X-ray images database.

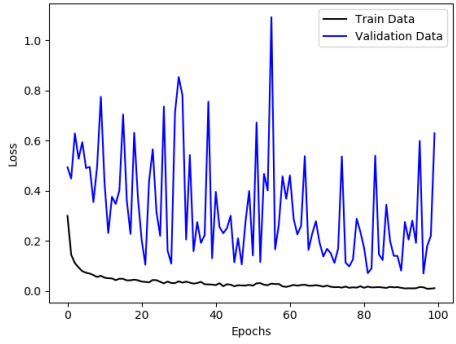
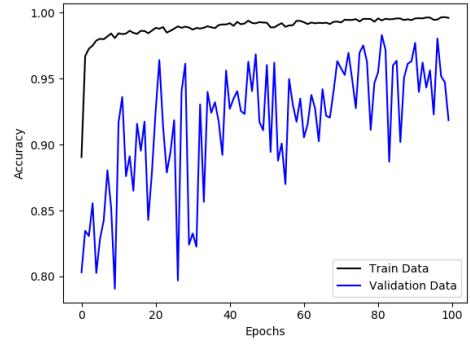

a) Loss diagram

b) Accuracy diagram

Figure 12: Performance metrics of CNN-sigmoid method using augmented COVID-19 X-ray images database: (a) loss function curve, and b) accuracy curve with 10-fold cross-validation strategy.

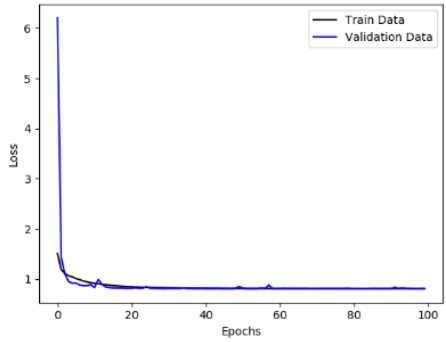
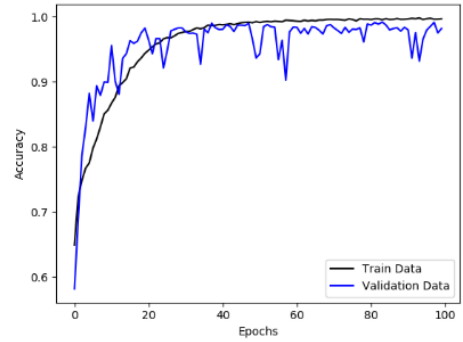

a) Loss diagram

b) Accuracy diagram

Figure 13: Performance metrics of CNN-SVM method using augmented COVID-19 X-ray images database: (a) loss function curve, and b) accuracy curve with 10-fold cross-validation strategy.

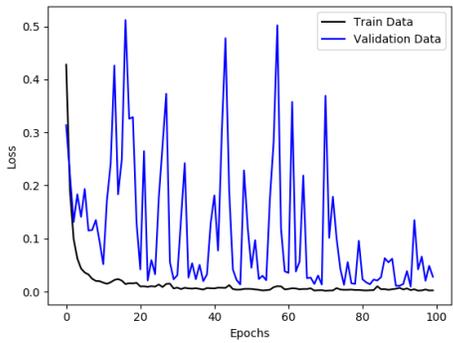
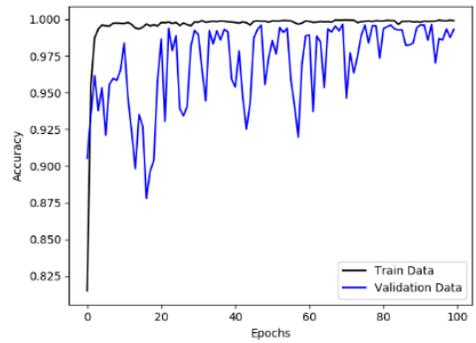

a) Loss diagram

b) Accuracy diagram

Figure 14: Performance metrics of CNN-sigmoid method with Sobel operator using augmented COVID-19 X-ray images database: (a) loss function curve, and b) accuracy curve with 10-fold cross-validation strategy.

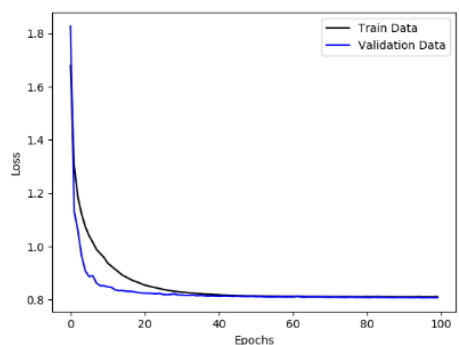 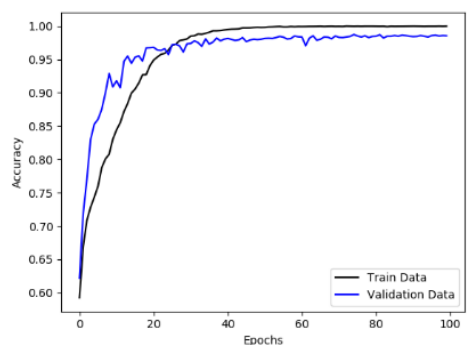

a) Loss diagram    b) Accuracy diagram

Figure 15: Performance metrics of CNN-SVM method with Sobel operator using augmented COVID-19 X-ray images database: (a) loss function curve, and b) accuracy curve with 10-fold cross-validation strategy.

Tables 3 and 4 depict the results obtained using various combination of networks with private database and augmented COVID-19 X-ray images database, respectively. Table 3 clearly shows the effect of using sigmoid or SVM classifiers and Sobel Filter when we used in our proposed method with our database.

Table 3. Various performance measures obtained using different combination of methods.

| Methods | Accuracy (%) | PPV (%) | Recall (%) | Specificity (%) | F1-score (%) | Loss | AUC |
|---|---|---|---|---|---|---|---|
| CNN-Sigmoid | 92.9418 | 98.00 | 92.90 | 91.13 | 95.00 | 0.2327 | 0.9203 |
| CNN-SVM | 98.2729 | 97.80 | 100 | 93.16 | 99.00 | 0.8088 | 0.9658 |
| CNN-Sigmoid +Sobel | 96.5435 | 97.50 | 98.30 | 90.42 | 97.80 | 0.1368 | 0.9438 |
| **CNN-SVM +Sobel** | **99.0248** | **98.70** | **100** | **95.23** | **99.40** | **0.8031** | **0.9770** |

Table 4 shows the evaluation performance measures obtained by applying different algorithms and combination of our methods using augmented COVID-19 X-ray images database.

Table 4. Evaluation performance measures obtained by applying different algorithms and combination of our methods using augmented COVID-19 X-ray images database.

| Methods | Accuracy (%) | PPV (%) | Recall (%) | Specificity (%) | F1-score (%) | Loss | AUC |
|---|---|---|---|---|---|---|---|
| Alqudah et al. (a) [47] | 99.46 | NA | 99.46 | 99.73 | NA | NA | NA |
| Alqudah et al. (b) [48] | 95.2 | 100 | 93.3 | 100 | NA | NA | NA |
| Haque et al. [49] | 99.00 | NA | NA | NA | NA | NA | NA |
| CNN-Sigmoid | 91.3883 | 93.40 | 94.00 | 89.96 | 92.20 | 0.6894 | 0.9192 |
| CNN-SVM | 98.2477 | 98.00 | 98.80 | 97.86 | 98.10 | 0.8044 | 0.9828 |
| CNN-Sigmoid +Sobel | 98.4636 | 98.80 | 98.40 | 98.68 | 98.20 | 0.0100 | 0.9848 |
| **CNN-SVM +Sobel** | 99.6156 | 99.60 | 99.80 | 99.56 | 99.50 | 0.8047 | 0.9968 |

It can be noted from Tables 3 and 4 that Sobel operator improved the performance of CNN-Sigmoid and CNN-SVM approaches in detecting COVID-19 significantly. Table 4 shows the results of our proposed method and other works on augmented COVID-19 X-ray images database. For better comparison between the achieved results in Tables 3 and 4, the results are also illustrated in Figures 16 and 17. They show the impact of using Sobel filtering in our algorithms. As SVM is a more robust classifier, when it is used in our algorithms, the performance has improved.

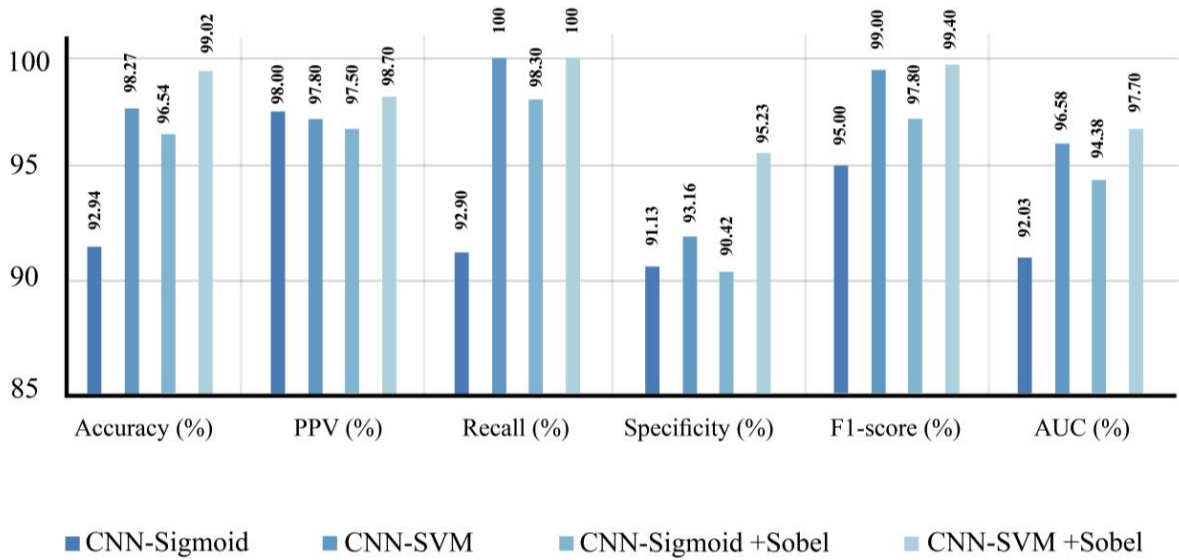

Figure. 16: Performance obtained using different methods with our private database for COVID-19 diagnosis.

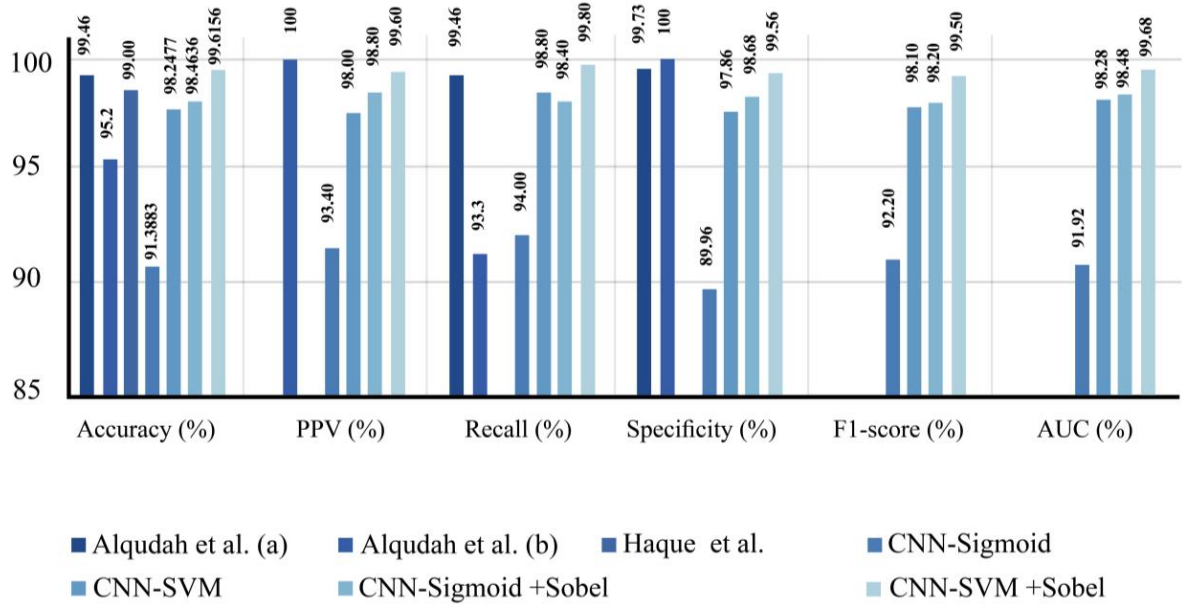

Figure. 17: Performance obtained using different methods with augmented COVID-19 X-ray images database for COVID-19 diagnosis.

Our proposed method is also tested with *six* public databases to evaluate the performance of our developed model. The public database can be accessed from this link: https://www.kaggle.com [50-55]. The details of the database and results obtained using our database are provided in Table 5. It can be noted from this table that using the Sobel filter can improve the performance of our algorithm. In all tests, using Sobel filter has a positive impact on the results. Also, CNN-SVM+Sobel performed better than others combinations. For all databases, CNN-Sigmoid+Sobel performed better than the rest of the combinations.

Table 5: Evaluation metrics obtained for our proposed method using different public databases.

| Database | Collected from | Number of cases | Method | Performance | | | | | | |
|---|---|---|---|---|---|---|---|---|---|---|
| | | | | Accuracy (%) | PPV (%) | Recall (%) | Specificity (%) | F1-score (%) | Loss | AUC |
| [50] | Bangladesh | 1820 | CNN-Sigmoid | 91.39 | 93.40 | 94.00 | 89.93 | 92.20 | 0.69 | 0.92 |
| | | | CNN-SVM | 98.25 | 98.00 | 98.80 | 97.87 | 98.10 | 0.80 | 0.98 |
| | | | CNN-Sigmoid +Sobel | 98.46 | 98.80 | 98.40 | 98.68 | 98.20 | 0.01 | 0.98 |
| | | | **CNN-SVM +Sobel** | 99.61 | 99.60 | 99.80 | 99.57 | 99.50 | 0.80 | 0.99 |
| [51] | India | 1160 | CNN-Sigmoid | 96.47 | 96.00 | 100 | 92.86 | 97.50 | 0.20 | 0.96 |
| | | | CNN-SVM | 97.82 | 97.10 | 100 | 95.46 | 98.30 | 0.80 | 0.98 |
| | | | CNN-Sigmoid +Sobel | 99.56 | 99.30 | 100 | 99.26 | 99.70 | 0.01 | 0.99 |
| | | | **CNN-SVM +Sobel** | 99.98 | 99.95 | 100 | 99.97 | 99.90 | 0.79 | 0.99 |
| [52] | Italy | 1550 | CNN-Sigmoid | 85.92 | 87.30 | 86.20 | 83.75 | 86.60 | 1.44 | 0.85 |
| | | | CNN-SVM | 86.60 | 97.30 | 87.30 | 70.99 | 89.50 | 0.84 | 0.84 |
| | | | CNN-Sigmoid +Sobel | 94.97 | 97.80 | 96.50 | 85.07 | 97.10 | 0.15 | 0.91 |
| | | | **CNN-SVM +Sobel** | 96.86 | 96.80 | 99.70 | 78.56 | 98.30 | 0.82 | 0.89 |
| [53] | India | 1120 | CNN-Sigmoid | 97.54 | 96.60 | 99.40 | 95.81 | 97.80 | 0.07 | 0.97 |
| | | | CNN-SVM | 99.10 | 99.50 | 98.80 | 99.37 | 99.20 | 0.80 | 0.99 |
| | | | CNN-Sigmoid +Sobel | 99.46 | 98.90 | 100 | 99.05 | 99.40 | 0.01 | 0.99 |
| | | | **CNN-SVM +Sobel** | 99.92 | 99.80 | 100 | 99.84 | 99.90 | 0.80 | 0.99 |
| [54] | Singapore | 460 | CNN-Sigmoid | 89.67 | 92.90 | 92.70 | 83.42 | 90.90 | 0.33 | 0.88 |
| | | | CNN-SVM | 97.61 | 99.70 | 96.50 | 99.33 | 97.90 | 0.80 | 0.98 |
| | | | CNN-Sigmoid +Sobel | 98.04 | 99.10 | 98.70 | 98 | 98.80 | 0.05 | 0.98 |
| | | | **CNN-SVM +Sobel** | 99.35 | 99.10 | 100 | 98 | 99.50 | 0.79 | 0.99 |
| [55] | Not available | 1930 | CNN-Sigmoid | 97.50 | 98.10 | 99.00 | 73.32 | 98.70 | 0.10 | 0.86 |
| | | | CNN-SVM | 97.30 | 97.90 | 99.30 | 66.69 | 98.50 | 0.82 | 0.83 |
| | | | **CNN-Sigmoid +Sobel** | 98.18 | 97.90 | 100 | 71.64 | 99.20 | 0.15 | 0.86 |
| | | | CNN-SVM +Sobel | 98.07 | 97.90 | 99.90 | 71.64 | 99.10 | 0.81 | 0.86 |

Meanwhile, in Table 6, the results of the proposed method applied on our database are compared with other researches who used different databases. Accordingly, the performance of our proposed method is better than other researches.

Table 6. Comparison of proposed CNN-SVM+Sobel method using private database with other methods in detecting COVID-19 using X-ray images from different private databases.

| Study | Number of Cases | Network | Train-Test | Evaluation Metrics |
|---|---|---|---|---|
| Hall et al. [15] | 455 images | VGG-16 and ResNet-50 | 10-fold | AUC: 0.997 |
| Hemdan et al. [18] | 50 images | DesnseNet, VGG16, MobileNet v2.0 etc. | 80-20% | F1 score: 91% |
| Abbas et al. [19] | 196 images | CNN with transfer learning | 70-30% | Accuracy: 95.12% Sensitivity: 97.91% Specificity: 91.87% PPV: 93.36% |
| Zhang et al. [21] | 213 images | ResNet, EfficientNet | 5-fold | Sensitivity: 71.70% AUC: 0.8361 |
| Narin et al. [24] | 100 images | ResNet50 | 10-fold | Accuracy: 98% |
| Ozturk et al. [34] | 625 images | Darknet-19 | 5-fold | Accuracy: 98.08% |
| Khan et al. [35] | 1251 images | CNN | 4-fold | Accuracy: 89.6% Sensitivity: 98.2% PPV: 93% |
| Iwendi et al. [56] | NA | Random Forest algorithm boosted by the AdaBoost algorithm | NA | Accuracy: 94% F1-score: 86% |
| Haghanifar et al. [37] | 780 images | DenseNet-121 U-Net | 75-25% | Accuracy: 87.21% |
| Oh et al. [38] | 502 images | DenseNet U-Net | 80-20% | Accuracy: 91.9 % |
| Tartaglione et al. [39] | 137 images | ResNet | 70-30% | Accuracy: 85% |
| Proposed Method | 1332 images | **CNN-SVM+Sobel** | 10-fold | Accuracy: 99.02% Sensitivity: 100% Specificity: 95.23% AUC: 0.9770 |

Figure 16 shows the performance obtained using different proposed methods with our private database for automated detection of COVID-19 patients using X-ray images. Figure 17 shows the performance obtained using various proposed methods with augmented COVID-19 X-ray images database for COVID-19 diagnosis. Figures 16 and 17 clearly show that our proposed CNN-SVM+Sobel model has performed better than rest of the methods on our database and augmented COVID-19 X-ray images database respectively. Our proposed method has performed better even using *six* public databases.

Advantages of our proposed method are as follows:

1. We collected a new database to validate our developed model.
2. Our proposed method is also tested on *six* public databases and showed excellent performance.
3. Data augmentation is used to enable it works with small databases.
4. Sobel filter is used to improve the performance of our method

Limitations of our proposed method are as follows:

1. Computational cost of different deep learning algorithm is high.
2. Limitation of input data is another weakness of our algorithm.

## 4. Conclusion

COVID-19 is currently one of the most life-threatening diseases endangering the health of many people globally. One of the main features of this disease is its rapid prevalence among people in the community. In this work, we have developed a novel COVID-19 detection system using X-ray images. In this work, we have used 333 X-ray images (77 COVID-19 + 256 normal) from Omid Hospital, Tehran to develop the model. First the images are subjected to Sobel filter to obtain the contours of the images and then fed to CNN model followed by SVM classifier. Our method is able to detect the COVID-19 cases correctly with an accuracy of 99.02%. The developed model has also yielded highest detection accuracy using *six* public databases. Hence, this justifies that our developed model is robust and accurate. In future, we intend to use this model to detect other chest related diseases like cancer, pneumonia, cystic fibrosis, infection, and chronic obstructive pulmonary disease (COPD).